\documentclass[prb,aps,twocolumn,preprintnumbers,10pt,floats,amsmath,showpacs,amssymb,superscriptaddress]{revtex4-1}

\usepackage{hyperref}
\usepackage{dcolumn}
\usepackage{bm}
\usepackage{amssymb}
\usepackage[latin1]{inputenc}
\usepackage{subfigure}
\usepackage{color}
\usepackage{float}
\usepackage{amsmath}
\usepackage{graphicx}
\usepackage{hyperref}
\usepackage{psfrag}
\usepackage{color}
\usepackage{simplewick}
\usepackage{appendix}
\usepackage{color}
\usepackage{comment}
\usepackage{bbold}
\usepackage{mathrsfs}

\allowdisplaybreaks

\definecolor{darkgreen}{rgb}{0.1,0.5,0.1}

\def \be{\begin{equation}}
\def \ee{\end{equation}} 

{\left\lbrace\begin{array}{@{}l@{}}}%
{\end{array}\right.}

\begin{document}

\title{Magnetic thermal switch for heat management at the nanoscale}

\author{Riccardo Bosisio}
\affiliation{SPIN-CNR, Via Dodecaneso 33, 16146 Genova, Italy}
\affiliation{NEST, Istituto Nanoscienze-CNR, and Scuola Normale Superiore, I-56126 Pisa, Italy}

\author{Stefano Valentini}
\affiliation{NEST, Scuola Normale Superiore, and Istituto Nanoscienze-CNR, I-56126 Pisa, Italy}

\author{Francesco Mazza}
\affiliation{NEST, Scuola Normale Superiore, and Istituto Nanoscienze-CNR, I-56126 Pisa, Italy}

\author{Giuliano Benenti}
\affiliation{Center for Nonlinear and Complex Systems, Universit\`a degli Studi dell'Insubria, 
		via Valleggio 11, 22100 Como, Italy}
\affiliation{Istituto Nazionale di Fisica Nucleare, Sezione di Milano, via Celoria 16, 20133 Milano, Italy}

\author{Rosario Fazio}
\affiliation{NEST, Scuola Normale Superiore, and Istituto Nanoscienze-CNR, I-56126 Pisa, Italy}

\author{Vittorio Giovannetti}
\affiliation{NEST, Scuola Normale Superiore, and Istituto Nanoscienze-CNR, I-56126 Pisa, Italy}

\author{Fabio Taddei}
\affiliation{NEST, Istituto Nanoscienze-CNR and Scuola Normale Superiore, I-56126 Pisa, Italy}

%------------------------------------------------------------------------------------------

\begin{abstract}

In a multi-terminal setup, when time-reversal symmetry is broken by a magnetic field, 
the heat flows can be managed by designing a device with programmable Boolean behavior.
We show that such a device can be used to implement operations, 
such as on/off switching,
reversal, selected splitting and swap of the heat currents.
For each feature, the switching from one working condition to the other 
is obtained by inverting the magnetic field.  
This offers interesting opportunities for conceiving a programmable setup, 
whose operation is controlled by an external parameter (the magnetic field) 
without need to alter voltage and thermal biases applied to the system.
Our results, generic within the framework of linear response, are illustrated
by means of a three-terminal electronic interferometer model.
\end{abstract}


\maketitle

%------------------------------------------------------------------------------------------
%
%------------------------------------------------------------------------------------------

\section{Introduction}
Heat management at the nanoscale is nowadays one of the leading research 
topics in many different scientific areas, including refrigeration and
thermometry~\cite{Giazotto2006}, coherent caloritronics~\cite{Perez2014},
thermoelectric energy 
conversion~\cite{Snyder2008,Shakouri2011,Dubi2011,Benenti2013,Jordan2013,Sanchez2013, Sothmann2014},
and information processing by utilizing phonons~\cite{Li2012}.
The overheating of microprocessor components is currently the most 
limiting factor in the development of information technology~\cite{Zhirnov2014},
which motivates the concern in finding alternative ways to control and 
evacuate heat in such devices.
Theoretical works led to the possibility of controlling the heat currents
and devise heat diodes~\cite{Terraneo2002} and transistors~\cite{Li2006}.
First experimental implementations exploiting 
phononic~\cite{Chang2006,Kobayashi2009,Tian2012}, 
electronic~\cite{Saira2007,Scheibner2008,Giazotto2014}, or 
photonic~\cite{Chen2014} thermal currents were also reported.

It has been shown that the presence of a magnetic field breaking time reversibility 
could in principle enhance the efficiency of thermoelectric 
devices~\cite{Benenti2011,Saito2011,Balachandran2013, Sanchez2014-1, Sanchez2014-2}. 
Interestingly a magnetic field allows for the simultaneous presence of reversible and
irreversible heat currents~\cite{Saito2013,Brandner2013}.
Indeed, in a generic multi-terminal setup, 
we can split the heat current
$J^{Q}_k$, flowing from the $k$-th terminal to the system,
into the sum  of a reversible and an irreversible part,
$J^{Q}_k=J^{Q(r)}_k+J^{Q(i)}_k$. Although the reversible component
changes sign by reversing the magnetic field $\bf{B}$, the irreversible
component is invariant under the inversion $\bf{B}\to-\bf{B}$.
Within the linear response regime, it can be  
shown~\cite{Saito2013,Brandner2013} that only the irreversible 
part of the current contributes to the entropy production.
On the other hand the reversible part vanishes for $\bf{B}=0$,
whereas for $\bf{B}\ne 0$ it becomes
arbitrarily large, giving rise, among other things, to the
possibility of dissipationless transport, i.e., to a thermal
machine operating at the Carnot efficiency with finite
power output~\cite{Benenti2011}.

In this paper we take advantage of the presence of reversible 
components of the heat currents to propose a magnetic thermal switch, 
a Boolean setup which allows the control of heat flow by making use of an 
external magnetic field as a selector of the working configuration.
For a generic multi-terminal device operating in the linear response 
regime, 
we show that by properly tuning the voltage biases we can access a 
broad spectrum of possible operating conditions, each of these being defined 
in terms of the behavior of the heat currents flowing through 
the system. 
Namely, it is possible to design
a \emph{programmable} device for the management of heat flows, 
allowing several Boolean features, such as selected splitting, on/off switching, 
reversal and swap of the heat currents. 
For each feature,
the magnetic field acts as a knob selecting one of the two possible 
working conditions, without the need to modify 
the reservoirs parameters (temperatures and electrochemical 
potentials): The switching from one working condition to the other
is obtained by inverting the direction of the magnetic field.\\
\indent A significant advantage of our approach is the absence of
temperature constraints: As long as the system operates in linear response,
our results hold.
In particular, the method we present is valid whether the heat
is transported by electrons, by phonons, or by both. Thus, remarkably, it constitutes a possible way of manipulating
phononic heat currents using a magnetic field. 
From a practical point of view, 
the implementation of our theoretical results would require a 
full characterization of the Onsager matrix, the major difficulty
being the measurement of the heat currents at the nanoscale, 
a challenge for which, however, important advances have been recently 
reported~\cite{Bourgeois,IBM}.
Moreover, assuming the system in contact with regions
having finite thermal capacitance rather than with ideally infinite reservoirs, 
the magnetic field switching could be used to control the temperatures of such regions,
allowing for instance the initialization of qubit states or the 
implementation of thermal logic gate operations~\cite{Li2012}.
We finally remark that there exists, in the literature, a variety of works on interferometer-based systems which, under broken time-reversal symmetry, would constitute natural physical realizations of our model (see, for instance, Refs.~\onlinecite{Ji2003, Neder2006, Neder2007, Roulleau2007, Giovannetti2008, Deviatov2008, Altimiras2010, Hofer2015, Sothmann2014, Sanchez2014-1, Sanchez2014-2}).

The paper is structured as follows: In Sec.~\ref{sec:thermalswitch} we 
describe the theoretical implementation of the magnetic thermal switch
for a general multi-terminal setup. 
Then in Sec.~\ref{sec:numerical} we present some results of numerical 
simulations using an interferometer in contact with three reservoirs as a toy model. Finally, we draw our 
conclusions in Sec.~\ref{sec:conclusions}. 
Details of the calculations and the derivation of the 
scattering matrix of the interferometer are gathered in the Appendices.

%------------------------------------------------------------------------------------------
%
%------------------------------------------------------------------------------------------

\section{Magnetic thermal switch}
\label{sec:thermalswitch}

In this section we discuss how a magnetic thermal switch can be implemented in a general multi-terminal setup. 
Let us consider a generic system in contact with $n$ reservoirs at temperatures $T_k = T +\Delta T_k$ and electrochemical potentials $\mu_k = \mu +\Delta \mu_k$, $T$ and $\mu$ being some equilibrium (reference) values. 
Let $\textbf{J}_k = (J^N_k,J^Q_k)$ denote the particle ($J^N_k$)
and heat ($J^Q_k$) currents from the $k$-th terminal to the system
and $\textbf{X}_k = (X^\mu_k,X^T_k) = (\Delta\mu_k/T,\Delta T_k/T^2)$
the conjugated affinities~\cite{Callen1985}. 
Within linear irreversible thermodynamics, the fluxes 
$\textbf{J}=(\textbf{J}_1,...,\textbf{J}_{n-1})^T$ 
and the conjugated affinities 
$\textbf{X}=(\textbf{X}_1,...,\textbf{X}_{n-1})^T$ 
are related as follows:
\begin{equation}
\textbf{J} = \textbf{L} \, \textbf{X},
\end{equation}
where $\textbf{L}$ is the Onsager matrix of kinetic 
coefficients~\cite{Callen1985} 
of dimension $2(n-1)\times 2(n-1)$. Note that, due to the constraints of 
particle and energy conservation, we can determine $\textbf{J}_n$ 
from the fluxes $\textbf{J}_1,...,\textbf{J}_{n-1}$. 
Moreover, we set the $n$-th reservoir as the reference one, with 
temperature $T$ and electrochemical potential $\mu$. 
In the presence of a magnetic field $\bf{B}$, time-reversal symmetry is
broken and the Onsager matrix $\bf{L}$ in general is not 
symmetric~\cite{Callen1985,Benenti2011,Saito2011}. 
The currents can be separated into 
\emph{reversible} components (which change sign by reversing 
$\bf{B}\to-\bf{B}$) and \emph{irreversible} components 
(which are invariant with respect to the inversion 
$\bf{B}\to-\bf{B}$)~\cite{Saito2013,Brandner2013}:
\begin{equation}
\textbf{J}^{(r)}\equiv\frac{\textbf{L}(\textbf{B})-\textbf{L}^T(\textbf{B})}{2}\,\textbf{X}, \quad \textbf{J}^{(i)}\equiv\frac{\textbf{L}(\textbf{B})+\textbf{L}^T(\textbf{B})}{2}\,\textbf{X}.
\label{eq:currentsrevirr}
\end{equation}
By virtue of the Onsager-Casimir relations
$L_{ij}(-\textbf{B})=L_{ji}(\textbf{B})$~\cite{Callen1985},
these currents have the properties that $\textbf{J}^{(r)}(\textbf{B})=-\textbf{J}^{(r)}(-\textbf{B})$ and $\textbf{J}^{(i)}(\textbf{B})=\textbf{J}^{(i)}(-\textbf{B})$.
In general these properties imply that $\textbf{J}(\textbf{B})=\textbf{J}^{(r)}(\textbf{B})+\textbf{J}^{(i)}(\textbf{B})\ne\textbf{J}^{(r)}(-\textbf{B})+\textbf{J}^{(i)}(-\textbf{B})=\textbf{J}(-\textbf{B})$.

The idea of the present proposal is to set  proper working conditions that enforce a given  target functional dependence between the thermal currents evaluated at $\textbf{B}$ and $-\textbf{B}$. For instance we may ask the current  $J^Q_k(\textbf{B})$ we get at the $k$-th contact, to be equal to twice  the current $J^Q_{k'}(-\textbf{B})$ one would get  at the $k'$-th contact when flipping the orientation of the  magnetic field. 
More generally, given a subset $K$ of the $n$ terminals of the system,  we will write our target functional dependence in the form of a linear constraint,
\begin{equation}
J^Q_k(-\textbf{B}) = \sum_{k'=1}^{n-1} \, x^{\text{(target)}}_{kk'} \, J^Q_{k'}(\textbf{B}), \qquad \forall k\in K\;,
\label{eq:JQm_JQp}
\end{equation}
where $x^{\text{(target)}}_{kk'}$ is an assigned $(n_0-1)\times (n-1)$ real matrix, with $n_0\leq n-1$ being the number of elements of $K$. 
This allows us to define different Boolean working conditions which, while
maintaining constant all the other system parameters, can be activated by simply operating on the relative orientation of the device with respect to the external magnetic field:
Special instances of these devices are explicitly discussed in the following subsections.

Once the Onsager matrix $L_{ij}(\textbf{B})$ and the coefficients $x^{\text{(target)}}_{kk'}$ are given, one can satisfy Eq.~\eqref{eq:JQm_JQp} 
by properly tuning the components of the affinity vector $\textbf{X}$. As a matter of fact, since
 the conditions~\eqref{eq:JQm_JQp}  are at most $n-1$ and the total  number of the affinity parameters is $2(n-1)$,
we can fulfill the former by only using half of the latter.
 In what follows we exploit this freedom to fix the values of the thermal affinity components $\{X_k^T\}$'s on each of the reservoirs~\cite{NOTE}, whereas using
 the $\{X_k^\mu\}$'s to  enforce the constraint~\eqref{eq:JQm_JQp}.
 When $n_0=n -1$, i.e., if we impose constraints on the  $J^Q_k(-\textbf{B})$'s of all the terminals of the system,
the procedure has the limitation of making the device operate \emph{only} for certain precise values of the currents flowing from each $k$-th reservoir. Indeed, imposing $n-1$ relations of the form of Eq.~\eqref{eq:JQm_JQp} univocally determines all the $\{X_k^\mu\}$'s and hence, assuming fixed temperatures, also all the $J^Q_k(\pm\textbf{B})$'s.  
This limitation is naturally overcome when $n_0$ is strictly smaller than $n-1$. 
For instance, one may choose to impose only one condition (i.e., $n_0=1$)  in order to 
leave all but one of the $\{X^\mu_k\}$'s unspecified. 
In particular, we could solve for $X^\mu_1$ to obtain:
\be \label{equcon}
X_1^\mu=a_2X_2^\mu + \ldots+a_{n-1}X^\mu_{n-1}+f(X_1^T,\ldots,X^T_{n-1}),
\ee
where $a_k$ are some functions of the Onsager matrix elements $L_{ij}$, 
whereas $f(X_1^T,\ldots,X^T_{n-1})$ depends on the temperatures and 
on the $L_{ij}$.
Setting $X_1^\mu=\text{const}$, Eq.~(\ref{equcon}) defines a ($n$-2)-dimensional 
hyper-surface in the space spanned by $(X_2^\mu,\ldots,X^\mu_{n-1})$.
Assuming constant temperatures, varying the electrochemical potentials 
along this surface allows changing the values of the heat currents, 
without compromising the working operation of the device. In this way, 
we use the extra degrees of freedom given by the reservoirs with free 
electrochemical potential to widen the operational range of the device to many 
values of the heat currents~\cite{Note1}.
%
%
%------------------------------------------------------------------------------------------
%
%
\subsection{Heat current multiplier}
In general, we may design a system in which the heat current in the $k$-th terminal becomes a fraction or a multiple of the original value when the magnetic field is reversed, which corresponds to having a diagonal matrix $x^{\text{(target)}}_{kk'}=\delta_{kk'} x_k$ in Eq.~\eqref{eq:JQm_JQp}:
$J^{Q}_k (-\textbf{B})=x_k J^{Q}_k (\textbf{B}) $.
Specifying a value for $x_k$ makes the system operate as a Boolean \emph{heat current multiplier} in which the two (Boolean) configurations correspond to an upward or a downward magnetic field. A illustration of such an operation for a three-terminal device is shown in Fig.~\ref{fig:operationalprinciple}(a).\\
%
%------------------------------------------------------------------------------------------
%
%
\vspace{-6mm}
\subsubsection{On/off switch}
Let us consider the specific case of a heat current multiplier in which $x_k = 0$: The device behaves as an \emph{on/off switch for the $k$-th terminal}, which means $J^{Q}_k(-\textbf{B})=J^{Q(r)}_k(-\textbf{B})+J^{Q(i)}_k(-\textbf{B})=0$, whereas $J^{Q}_k(\textbf{B})=J^{Q(r)}_k(\textbf{B})+J^{Q(i)}_k(\textbf{B})\ne0$. It is then clear that $J^{Q(i)}_k(\textbf{B})$ and $J^{Q(r)}_k(\textbf{B})$ have the same magnitude and the same sign and add up giving a finite current, whereas the two terms cancel out upon magnetic field reversal, resulting in a \emph{vanishing} heat current.
This principle could be used, for instance, to implement a $n$-terminal \emph{selector for the heat path} in which an upward magnetic field allows the flow of heat through $l$ channels while blocking it into the remaining $(n-l)$ ones, and vice-versa by reversing $\textbf{B}\to -\textbf{B}$. A schematic of such an operation for a three-terminal device is shown in Fig.~\ref{fig:operationalprinciple}(b).
%
%------------------------------------------------------------------------------------------
%
\subsubsection{Fully reversible heat}
Another interesting configuration is obtained by setting $x_k = -1$  in which case the heat current is \emph{fully reversible}~\cite{note_reversibility} ($J^{Q(i)}_k=0$).
As an application, one could conceive a device in which the heat currents flowing through some (or all) the channels simultaneously flip their sign upon reversing the magnetic field. This, among other things, would offer the possibility of switching from a ``refrigerator'' mode for a specific reservoir to a ``thermal engine'' one by simply using the external magnetic field, without needing to modify the gradients in the reservoirs. A schematic of such an operation for a three-terminal device is shown in Fig.~\ref{fig:operationalprinciple}(c). Note that, by analogous considerations, $x_k = 1$ corresponds to the case of \emph{fully irreversible} heat currents, which is however much less interesting because in this situation reversing the magnetic field has no effect.
%
%------------------------------------------------------------------------------------------
%
%
\subsection{Heat current swap}
\begin{figure}[h!]
\centering
\includegraphics[width=0.84\columnwidth,  keepaspectratio]{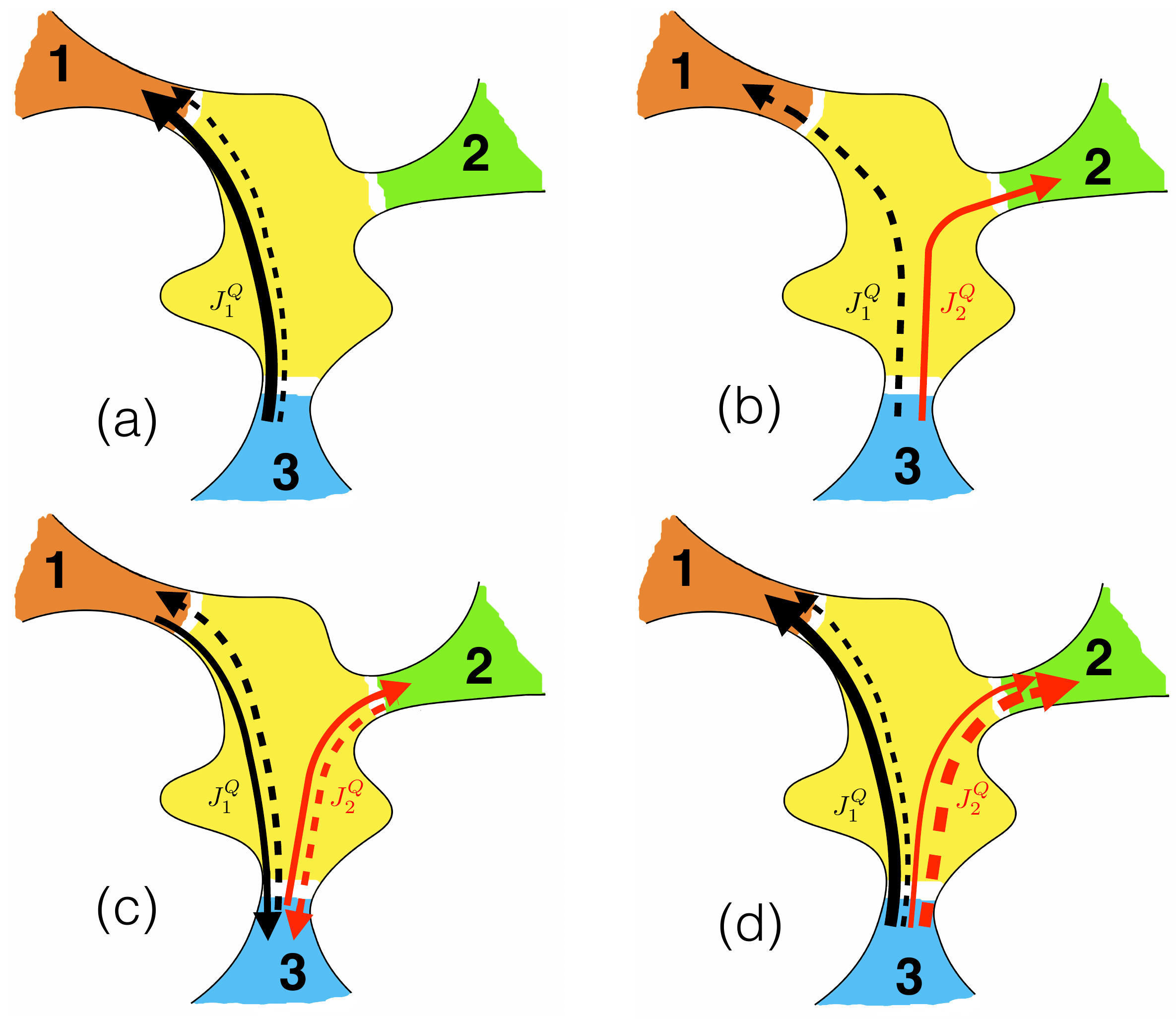}
\caption{Examples of operational principles for a three-terminal magnetic thermal switch. The different panels illustrate the heat-current (a) multiplier,  (b) selector, (c) reversal and (d) swap configurations, respectively. The working operation is selected by choosing either $+\bf{B}$ or $-\bf{B}$. Solid[dashed] lines correspond to $J^Q(+{\bf B})$[$J^Q(-{\bf B})$], whereas black(red) lines refer to currents flowing from terminal 1(2). Notice that in panel (a) lines of different thicknesses have been used to emphasize the increase/decrease of the heat currents magnitude before and after the magnetic field reversal.}
\label{fig:operationalprinciple}
\end{figure}

The matrix $x^{\text{(target)}}_{kk'}$ which defines our target~(\ref{eq:JQm_JQp}) does not need to be diagonal. For instance let us consider the case where 
$x^{\text{(target)}}_{kk'}=x^{\text{(target)}}_{k'k}=1$ and $x^{\text{(target)}}_{kk}=x^{\text{(target)}}_{k'k'}=0$, which implements a \emph{heat current swap} between reservoirs $k$ and $k'$. This configuration couples heat currents flowing from different terminals, whereas in the previous ones the conditions were imposed on each single reservoir independently. Such a choice for $x^{\text{(target)}}_{kk'}$ results in having $J^{Q}_k(\textbf{B})=J^{Q}_{k'}(-\textbf{B})$ and $J^{Q}_{k'}(\textbf{B})=J^{Q}_k(-\textbf{B})$, i.e., the two heat currents are swapped by reversing the magnetic field, as pictorially shown in Fig.~\ref{fig:operationalprinciple}(d) for a three-terminal case. Besides, we notice that in this situation the reversible and irreversible components of the heat currents satisfy the conditions: $J_k^{Q(i)}=J_{k'}^{Q(i)}$ and $J_k^{Q(r)}=-J_{k'}^{Q(r)}$.

It is worth stressing that in a generic multi-terminal setup different working conditions can co-exist: For instance, some channels can be configured as heat current selectors, whereas others may operate as multipliers, make heat reversal or swap.
\begin{figure}[t!]
\centering
\includegraphics[width=0.72\columnwidth,  keepaspectratio]{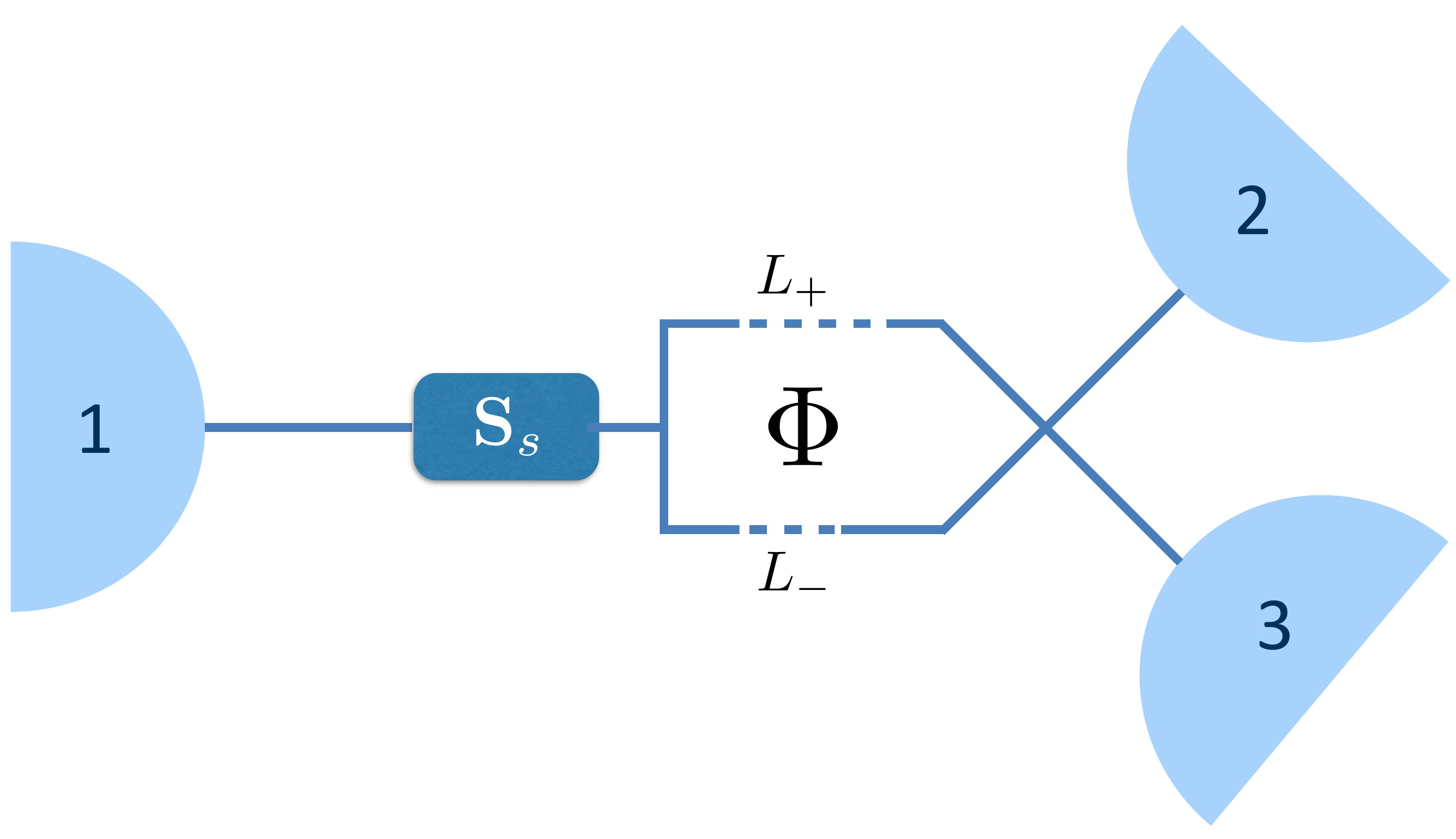}
\caption{Sketch of the three-terminal magnetic thermal switch studied numerically: an electronic interferometer, pierced by a magnetic flux $\Phi$ and in contact with three reservoirs at different temperatures $T_k$ and electrochemical potentials $\mu_k$ ($k=1,2,3$). The scattering region $S_s$ inside channel 1 breaks the particle-hole symmetry. $L_+$ and $L_-$ are the interference paths and must be different in order to observe interference at the end of the device.} 
\label{fig:interferometer}
\end{figure}
\vspace{-6mm}
%
%
%------------------------------------------------------------------------------------------
%
%------------------------------------------------------------------------------------------
%
%
\section{Simple model}
\label{sec:numerical}

In order to illustrate the effects discussed in the previous section, we study a simple noninteracting model consisting of a three-terminal interferometer sketched in Fig.~\ref{fig:interferometer}. We assume for simplicity low temperatures, so that electrons are the only heat carriers.
Under these conditions, the electronic transport through the device is coherent, which allows us to follow a scattering approach\cite{Datta:1995}.
The system consists of an interference loop, for example, made by two clean wires and connected to three electronic reservoirs with temperatures $T_k$ and electrochemical potentials $\mu_k$ ($k=1,2,3$). 
A magnetic field $\textbf{B}$ orthogonal to the interferometer plane generates a magnetic flux $\Phi$ piercing the loop, which will be the relevant parameter in the following discussion (from here on, we will assume $\Phi$ to be expressed in units of $h/2e$).
A scattering region $S_s$ is inserted into channel 1, having the effect of breaking the particle-hole symmetry $E \rightarrow -E$ (having set $\mu=0$ as the reference zero energy) in order to have finite non-diagonal Onsager coefficients. 
The specific choice of such scatterer is not important for the present discussion, as it does not alter the results at a qualitative level.
Further details on the computation of the scattering matrix of this system are given in Appendix~\ref{app1}.
Following the notation of the previous section, we set the reservoir 3 as the reference one ($\{\mu_3,T_3\}\equiv \{\mu,T\}$), and we express the particle and heat currents flowing from the other two reservoirs via the following $4\times 4$ linear system~\cite{Mazza2014}:
\begin{equation}
\begin{pmatrix}
J_1^N \\
J_1^Q \\
J_2^N \\
J_2^Q \\
\end{pmatrix}
=
\begin{pmatrix}
L_{11} & L_{12} & L_{13} & L_{14} \\
L_{21} & L_{22} & L_{23} & L_{24} \\
L_{31} & L_{32} & L_{33} & L_{34} \\
L_{41} & L_{42} & L_{43} & L_{44} \\
\end{pmatrix} 
\begin{pmatrix}
X_1^\mu \\
X_1^T \\
X_2^\mu \\
X_2^T \\
\end{pmatrix}.
\label{omatrix}
\end{equation}
The coefficients $L_{ij}$ are functions of the magnetic flux $\Phi$ and therefore of the applied magnetic field $\textbf{B}$.
Their explicit expressions are given by Eqs.~\eqref{eq:onsagers_app}, derived in Appendix~\ref{app2}.
The reversible ($r$) and irreversible ($i$) components of the heat currents  $J^Q_1$ and $J^Q_2$ are as follows~\cite{note_nointeracting}:
\begin{align}
&J^{Q(r)}_1 = \frac{L_{23}-L_{32}}{2} X^\mu_2 + \frac{L_{24}-L_{42}}{2} X^T_2, \cr
&J^{Q(i)}_1 = L_{21}X^\mu_1 + L_{22}X^T_1 + \frac{L_{23}+L_{32}}{2} X^\mu_2 + \frac{L_{24}+L_{42}}{2} X^T_2, \cr
&J^{Q(r)}_2 = \frac{L_{41}-L_{14}}{2} X^\mu_1 + \frac{L_{42}-L_{24}}{2} X^T_1, \cr
&J^{Q(i)}_2 = \frac{L_{41}+L_{14}}{2} X^\mu_1 + \frac{L_{42}+L_{24}}{2} X^T_1 + L_{43}X^\mu_2 + L_{44}X^T_2.\cr
\label{eq:heatcurrents_3t}
\end{align}
\begin{figure}[H]
\centering
\includegraphics[width=0.95\columnwidth]{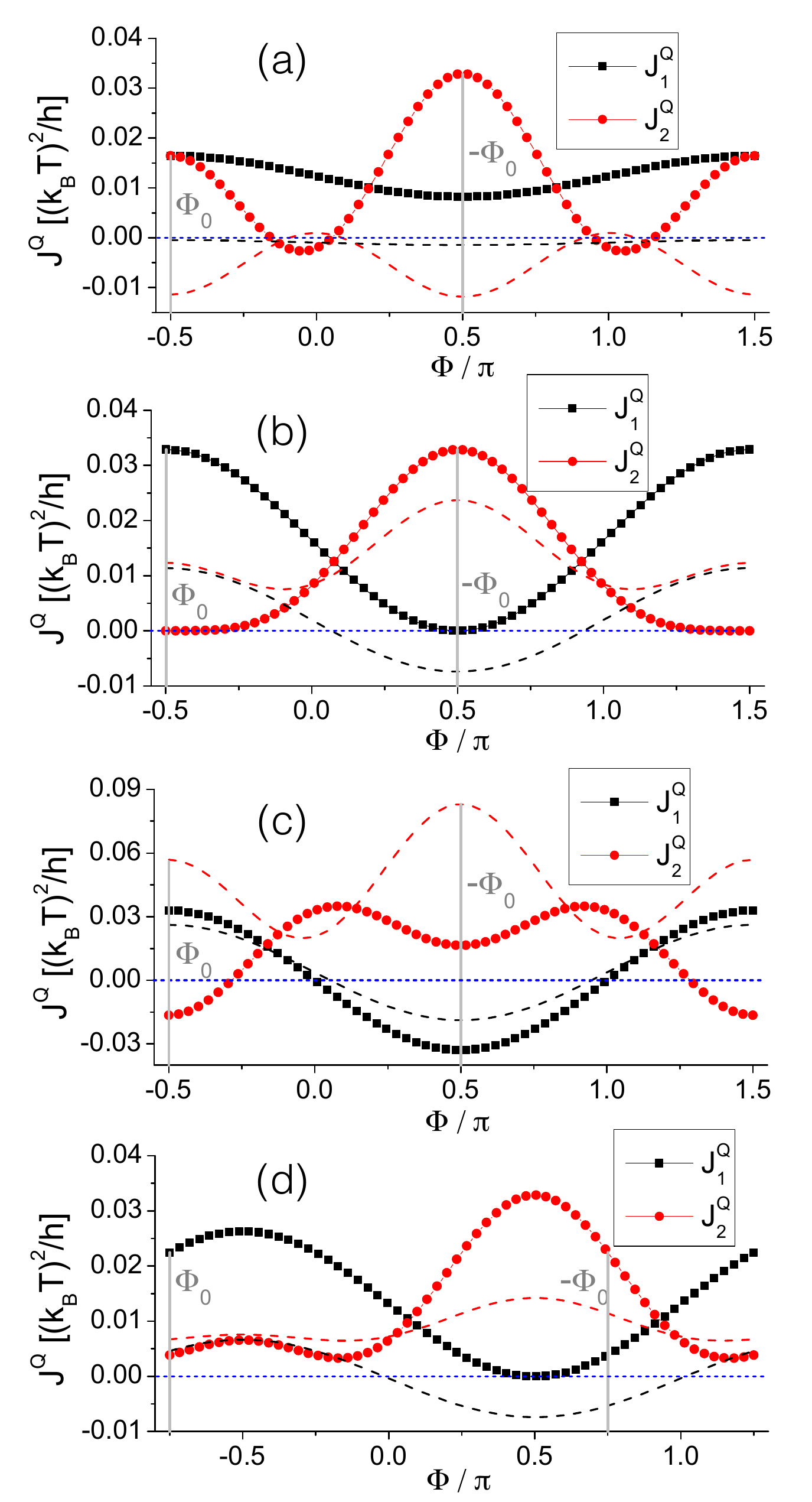}
\caption{Working operations of the three-terminal magnetic thermal switch discussed in the text. The heat currents through channels 1 (black squares) and 2 (red circles) are shown as a function of the magnetic flux $\Phi$ enclosed in the interferometer. For completeness, the particle currents (black and red dashed lines) are also shown, to emphasize that they do not follow the same behaviors as the heat currents. (a) \textit{Heat current multiplier}: By reversing $\Phi$ from $\Phi_0=-\pi/2$ to $-\Phi_0=+\pi/2$, $J^Q_1$ is halved whereas $J^Q_2$ is doubled. (b) \textit{Heat path selector}: For $\Phi_0=-\pi/2$, $J^Q_1$ is finite whereas $J^Q_2$ is blocked. The situation is opposite by reversing $\Phi$ to $-\Phi_0$. (c) \textit{Heat current reversal}: By reversing $\Phi$ from $-\pi/2$ to $+\pi/2$, the signs of both $J^Q_1$ and $J^Q_2$ flip. (d) \textit{Heat current swap}: By reversing $\Phi$ from $-3\pi/4$ to $+3\pi/4$, the values of $J^Q_1$ and $J^Q_2$ are interchanged. The parameters are $k_BT=1$, $\mu=0$, $X_1^T=0.025$, $X_2^T=0.01$ [except in (c), where $X_1^T=-0.005$ and $X_2^T=0.005$] and the difference between the interference paths in the upper/lower interferometer arms is $\Delta (kL)\equiv k(L_+-L_-)=\pi/2$. Dotted blue lines are guides to the eye at $J^Q=0$, whereas gray lines highlight the magnetic flux values $\Phi=\pm\Phi_0$ selecting the two Boolean configurations.}
\label{fig:results}
\end{figure}
Once the $L_{ij}$ coefficients \emph{for a given magnetic flux} $\Phi_0$ are calculated, for fixed $X_{1,2}^T$ different Boolean working conditions can be achieved by tuning the electrochemical potentials (and hence $X_{1,2}^\mu$) in order to impose Eq.~\eqref{eq:JQm_JQp} in both channels 1 and 2. Then, the switch is realized by reversing the magnetic field $\textbf{B}\to -\textbf{B}$, and hence the flux $\Phi_0 \to -\Phi_0$.
In order to illustrate the effects outlined in the previous section, we focus here below on the same four working conditions, by properly choosing the values of $x^{\text{(target)}}_{kk'}$ appearing in Eq.~\eqref{eq:JQm_JQp}. The numerical results are summarized in Fig.~\ref{fig:results}: Notice that both the heat (symbols) and the particle (dashed lines) currents are shown, to stress that they are not constrained to follow the same behaviors.
\begin{itemize}
	\item \textit{Heat current multiplier}, $(x_1,x_2)=(1/2,2)$. %: this implies to choose $(a_1,a_2)=(3,-3)$. 
	In this case the heat currents satisfy:
	\begin{align}
	& J^Q_1(-\textbf{B}) = \frac{1}{2}\,J^Q_1(+\textbf{B}),\cr
	& J^Q_2(-\textbf{B}) = 2\,J^Q_2(+\textbf{B}),
	\end{align}
  that is, by reversing the magnetic field $J^Q_1$ is halved whereas $J^Q_2$ is doubled. Under these conditions, by using Eqs. \eqref{eq:JQm_JQp} and \eqref{eq:heatcurrents_3t}, it is straightforward to see that the reversible and irreversible components of the heat currents are related via: $J^{Q(i)}_1(\textbf{B})=3J^{Q(r)}_1(\textbf{B})$ and $J^{Q(i)}_2(\textbf{B})=-3J^{Q(r)}_2(\textbf{B})$.
	The behavior of the heat currents flowing through the interferometer as a function of the magnetic flux in this configuration is shown in Fig.~\ref{fig:results}(a) for the interferometer described above.
	\item \textit{Heat path selector}, $(x_1,1/x_2)=(0,0)$. %: this implies to choose $(a_1,a_2)=(1,-1)$, by virtue of Eq.~\eqref{eq:alphavsa}. 
	In this case the heat currents satisfy $J^{Q}_1(\textbf{B}) \neq 0 $, $J^Q_2(\textbf{B}) = 0$ and $J^{Q}_1(-\textbf{B}) = 0 $, $J^Q_2(-\textbf{B}) \neq 0$, i.e., for an upward magnetic field, heat transfer is allowed between the system and reservoir 1, while being blocked between the system and reservoir 2. This situation is reversed by changing $\textbf{B}\to -\textbf{B}$ ($\Phi_0\to -\Phi_0$). Furthermore, according to Eqs. \eqref{eq:JQm_JQp} and \eqref{eq:heatcurrents_3t}, the reversible and irreversible components of the heat currents are related via: $J^{Q(i)}_1(\textbf{B})=J^{Q(r)}_1(\textbf{B})$ and $J^{Q(i)}_2(\textbf{B})=-J^{Q(r)}_2(\textbf{B})$. The behavior of $J^Q_1$ and $J^Q_2$ is shown in Fig.~\ref{fig:results}(b).
	\item \textit{Heat current reversal}, $(x_1,x_2)=(-1,-1)$. %: this implies to choose $(a_1,a_2)=(0,0)$. 
	In this case the heat currents are purely reversible, that is, $J^{Q(i)}_1 = J^{Q(i)}_2 = 0$. Reversing the magnetic flux through the interferometer makes them simultaneously change their sign. The behavior of $J^Q_1$ and $J^Q_2$ is shown in Fig.~\ref{fig:results}(c). Note that at $\Phi_0=-\pi/2$ both $J^Q_1$ (black squares) and $J^N_1$ (black dashed line) are positive. This, together with the fact that $X^\mu_1>0$ in this case, means that the system is acting as a \emph{local refrigerator}\cite{Imry2014} for the reservoir 1, exploiting a positive $\Delta\mu_1$ to extract heat from a cold bath ($X^T_1<0$). Conversely, at $\Phi=-\Phi_0=\pi/2$, both $J^Q_1$ and $J^N_1$ have changed their sign: the system is now performing work driving particles against $\Delta\mu_1$, thus operating as a thermal engine. Notice that the same reasoning does not hold for reservoir 2 in which, upon reversing the magnetic flux, the sign of $J^Q_2$ flips whereas that of $J^N_2$ does not.
	\item \textit{Heat current swap}, $(x_{12},x_{21})=(1,1)$. 
	The heat currents satisfy $J^{Q}_1(\textbf{B})=J^{Q}_{2}(-\textbf{B})$ and $J^{Q}_{2}(\textbf{B})=J^{Q}_1(-\textbf{B})$, that is, the two heat currents are swapped by reversing the magnetic field. Furthermore, according to Eqs. \eqref{eq:JQm_JQp} and \eqref{eq:heatcurrents_3t}, the reversible and irreversible components of the heat currents are related via: $J_1^{Q(i)}(\textbf{B})=J_2^{Q(i)}(\textbf{B})$ and $J_1^{Q(r)}(\textbf{B})=-J_2^{Q(r)}(\textbf{B})$. The behavior of $J^Q_1$ and $J^Q_2$ is shown in Fig.~\ref{fig:results}(d).
\end{itemize}
%

%------------------------------------------------------------------------------------------
%
%------------------------------------------------------------------------------------------

\section{Conclusions}
\label{sec:conclusions}

In this article we have shown that a magnetic thermal switch can
be implemented within the framework of linear response, taking 
advantage of the generic existence of reversible heat currents when 
time reversal symmetry is broken. Such a device could allow the 
implementation of several Boolean features, such as on/off switching,
reversal, selected splitting, and swap of the heat currents.  
For each feature, the switching from one working condition to 
the other is obtained by inverting the direction of an 
applied magnetic field. 
Quite interestingly, it is possible to change the operating mode of the
device (from a power generator to a refrigerator) with respect to one of the reservoirs by 
inverting the external driving parameter, i.e., the magnetic field,
at fixed electrochemical potentials and temperatures of the
reservoirs.

A further advantage of our magnetic switch would arise in the perspective of conceiving
 a more complex programmable system, made of (for instance) an array of $N$ simpler subsystems.
These may be set up to operate in a variety of independent configurations, 
but always defined in terms of conditions of the form Eq.~\eqref{eq:JQm_JQp}. 
We could imagine designing an array of $N$ elements that are all initialized
 in the same state (say, for upward magnetic field $\textbf{B}$), but that upon reversing 
$\textbf{B} \to -\textbf{B}$ go to (possibly all different) final states.
We stress once more that acting on a \emph{single} parameter - the magnetic field - would be enough
 to achieve this operation and to reinitialize them in a subsequent moment, if needed.

Note that, although we have illustrated the magnetic thermal switch
for a low-temperature interferometer model, with the heat carried 
by the electrons, the mechanism discussed in this paper is generic
for any system with the time-reversal symmetry broken by a magnetic 
field. A magnetic thermal switch could be in principle 
implemented also when both fermionic and bosonic reservoirs are 
present. Indeed, as shown in Ref.~\onlinecite{Aharonyphonons}
due to the electron-phonon coupling the Onsager kinetic coefficients connecting 
the phononic heat currents from the bosonic reservoirs to the affinities 
for the fermionic terminals, in general are not even functions of 
the magnetic field. As a consequence, the phononic heat current 
generally exhibits a reversible component, and our theory can be applied.

\section*{Acknowledgements}

Stimulating discussions with F. Giazotto are gratefully acknowledged.
This work has been supported by Grant No. MIUR-FIRB2013 -- Project Coca (Grant No.~RBFR1379UX), by the EU project ``ThermiQ'', by MIUR-PRIN ``Collective quantum phenomena: from strongly correlated systems to quantum simulators'', by the EU project COST Action Project No. MP1209 ``Thermodynamics in the quantum regime'' and by the by the EU project COST Action Project No. MP1201 ``Nanoscale
Superconductivity: Novel Functionalities through Optimized Confinement of Condensate and Fields''.

%------------------------------------------------------------------------------------------
%
%------------------------------------------------------------------------------------------

\appendix

\section{Modeling the interferometer}\label{app1}

\begin{figure}[!h]
\centering
\includegraphics[width=\columnwidth,  keepaspectratio]{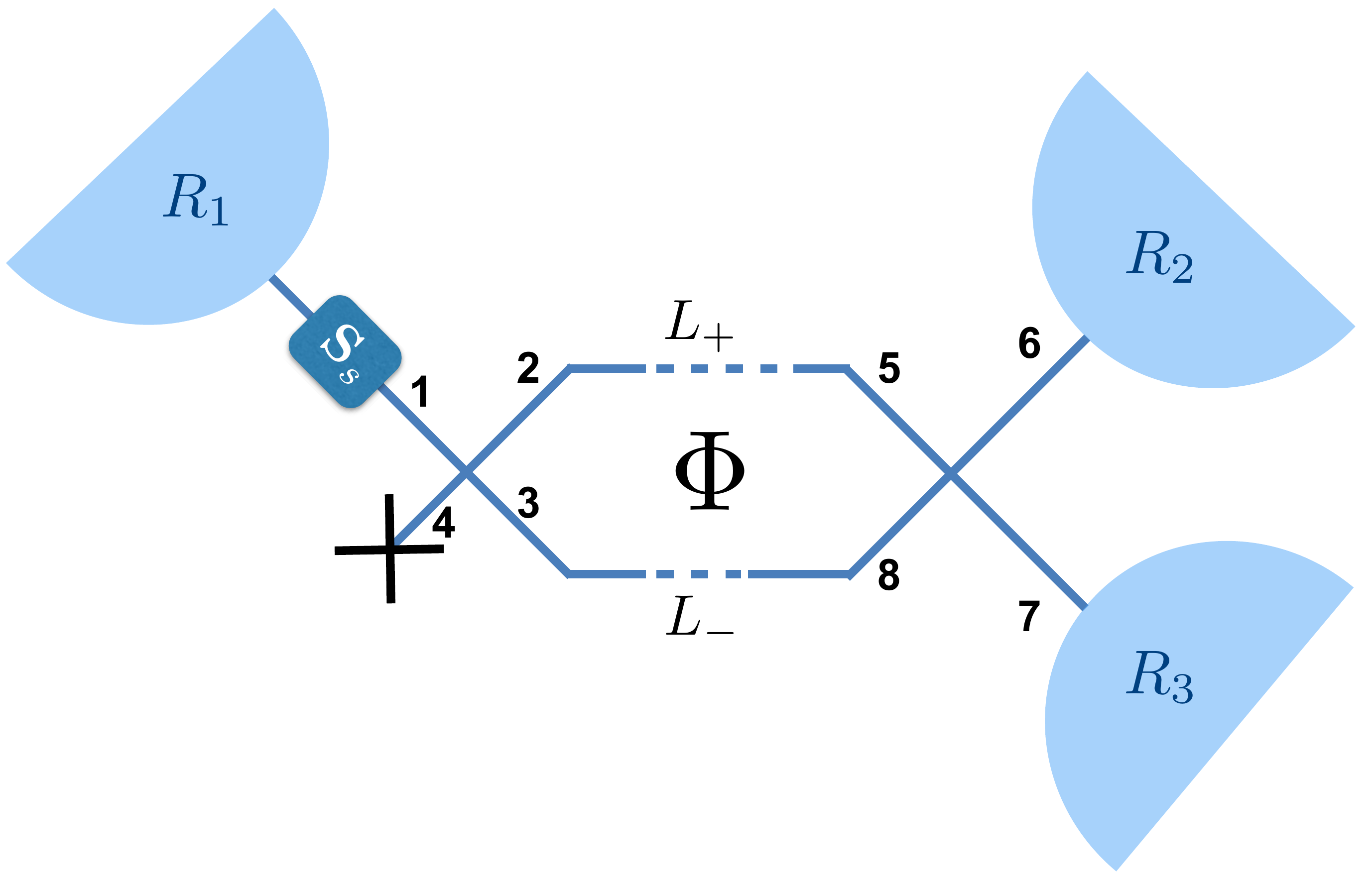}
\caption{Sketch of the system used to model the interferometer discussed in Sec. \ref{sec:numerical}. Two identical four-arm beam splitters are connected via two clean electronic waveguides of lengths $L_+$ and $L_-$, forming an interference loop which is pierced by a magnetic flux $\Phi$. A scatterer $S_s$ is inserted into arm 1 in order to break the particle-hole symmetry. The numbers from 1 to 8 refer to the arms of the beam splitters and label the transmission and reflection amplitudes $t_{ij}$ and $r_{ij}$ (in particular, channel 4 is assumed to be totally reflective). The system is connected to three electronic reservoirs $R_1$, $R_2$, and $R_3$.}
\label{fig:figura_sistema_appendice}
\end{figure}

In this section we outline the procedure followed to compute the scattering matrix of our interferometric system.
We start by considering an interferometer realized by connecting two four-arms beam-splitters via two clean wires (see Fig.~\ref{fig:figura_sistema_appendice}). 
For simplicity, we assume the beam splitters to be identical and symmetric, that means, each one is described by a scattering matrix of the form:
\begin{align}
S_{bs}&=
\begin{pmatrix}
r_{11} & t_{12} & t_{13} & t_{14} \\
t_{21} & r_{22} & t_{23} & t_{24} \\
t_{31} & t_{32} & r_{33} & t_{34} \\
t_{41} & t_{42} & t_{43} & r_{44} \\
\end{pmatrix}
=\cr
&=
\begin{pmatrix}
0 & 1/\sqrt{2} & 1/\sqrt{2} & 0 \\
1/\sqrt{2} & 0 & 0 & 1/\sqrt{2} \\
1/\sqrt{2} & 0 & 0 & -1/\sqrt{2} \\
0 & 1/\sqrt{2} & -1/\sqrt{2} & 0 \\
\end{pmatrix}.
\label{BS_scattering}
\end{align}
The matrix $S_{bs}$ describes a 50:50 beam splitter of electron waves, for which all the reflection terms are zero, and such that particles entering through one arm can be transmitted into two of the other three, with equal probability one half.
According to the notation of Fig.~\ref{fig:figura_sistema_appendice}, we have to compose the scattering matrices of the two individual beam splitters with the free propagation phase terms associated with the two interference paths. These terms are products of both the geometric (Aharonov-Bohm) phase and the dynamical phase exponentials:
\begin{align}
&f_{25}=f_g^{+}\times f_d^{+},\quad f_{38}=(f_g^{-})^*\times f_d^{-},\cr
&f_{52}=(f_g^{+})^*\times f_d^{+},\quad f_{83}=f_g^{-}\times f_d^{-},
\label{eq:phases}
\end{align}
where $f_g^+\,f_g^-=\text{exp} \{ i \Phi \}$, $f_d^{\pm}=\text{exp}\{i kL_{\pm}\}$, the $\pm$ signs denote the upper (+) and lower (-) interference arms of lengths $L_{\pm}$, the complex conjugation accounts for the electron traveling direction, $\Phi$ denotes the magnetic flux enclosed in the interferometer and $k$ is the Fermi wavevector.
For simplicity we neglect the energy dependence of the free-propagations.
The resulting scattering matrix describes the propagation among channels 1,4,6 and 7, and reads
\begin{equation}
S^{(1)}_{i}=
\begin{pmatrix}
0 & 0 & t_{16} & t_{17} \\
0 & 0 & t_{46} & t_{47} \\
t_{61} & t_{64} & 0 & 0 \\
t_{71} & t_{74} & 0 & 0 \\
\end{pmatrix},
\label{MZ_scattering_1}
\end{equation}
where the various coefficients $t_{pq}$ account for the different possible paths along which particles can travel from $p$ to $q$,
\begin{align}
& t_{16}=t_{12} f_{25} t_{56} + t_{13} f_{38} t_{86}, \quad  t_{17}=t_{12} f_{25} t_{57} + t_{13} f_{38} t_{87},\cr
& t_{46}=t_{42} f_{25} t_{56} + t_{43} f_{38} t_{86}, \quad  t_{47}=t_{42} f_{25} t_{57} + t_{43} f_{38} t_{87},\cr
& t_{61}=t_{65} f_{52} t_{21} + t_{68} f_{83} t_{31}, \quad  t_{71}=t_{75} f_{52} t_{21} + t_{78} f_{83} t_{31},\cr
& t_{64}=t_{65} f_{52} t_{24} + t_{68} f_{83} t_{34}, \quad  t_{74}=t_{75} f_{52} t_{24} + t_{78} f_{83} t_{34}.
\end{align}
Now, since we are interested in a three-terminal configuration, we impose that one of the channels (say, channel 4) behaves as a purely reflective mirror characterized by a reflection amplitude $r=-1$. The interferometer scattering matrix thus reduces to a $3\times3$ matrix,
\begin{equation}
S^{(2)}_{i}=
\begin{pmatrix}
r'_{11} & t'_{16} & t'_{17}  \\
t'_{61} & r'_{66} & t'_{67}  \\
t'_{71} & t'_{76} & r'_{77}  \\
\end{pmatrix}
=
\begin{pmatrix}
0  & t_{16} & t_{17} \\
t_{61} & t_{64}\, r\, t_{46} & t_{64}\, r\, t_{47} \\
t_{71} & t_{74}\, r\, t_{46} & t_{74}\, r\, t_{47} \\
\end{pmatrix}.
\label{MZ_scattering_2}
\end{equation}
Finally, in order to break the particle-hole symmetry, we insert in channel 1 an energy-dependent scattering region, described by a scattering matrix
\begin{equation}
S_{s}=
\begin{pmatrix}
\rho & i\tau   \\
i\tau & \rho   \\
\end{pmatrix},
\label{S_scattering}
\end{equation}
where $\rho, \tau \ge 0$ and 
such that

\begin{equation}
\tau = 
\begin{cases}
\label{eq:scatterer}
%\tau=1\qquad\text{if}\qquad E>0 \cr \tau=0\qquad\text{elsewhere},
1,\qquad\text{if}\qquad E>0 \cr 0,\qquad\text{elsewhere},
\end{cases}
\end{equation}
with $\rho^2=1-\tau^2$. This energy step would naturally be implemented using a well tuned electronic constriction, such as a quantum point contact\cite{vanHouten1992}.
The final expression for the scattering matrix of the whole system is as follows:
\begin{align}
S&=
\begin{pmatrix}
r''_{11} & t''_{16} & t''_{17}  \\
t''_{61} & r''_{66} & t''_{67}  \\
t''_{71} & t''_{76} & r''_{77}  \\
\end{pmatrix}
=\cr
&=
\begin{pmatrix}
\rho  & i\tau\,t'_{16} & i\tau\,t'_{17} \\
i\tau\,t'_{61} & r'_{66}+ t'_{61}\, \rho\, t'_{16} & t'_{67}+t_{61}\, \rho\, t'_{17} \\
i\tau t'_{71} & t'_{76}+t'_{71}\, \rho\, t'_{16} & r'_{77}+t_{71}\, \rho\, t'_{17} \\
\end{pmatrix}.
\label{scattering}
\end{align}
It is worth observing that, having initialized all the $r_{pq}$ and $t_{pq}$ in Eq.\eqref{BS_scattering}, the remaining (relevant) free parameters in the scattering matrix above are the difference between the paths in the upper/lower interference arms, $\Delta L=L_+-L_-$, and the magnetic flux enclosed in the interferometer loop, $\Phi$ [see Eq.\eqref{eq:phases}].
\section{Calculation of the Onsager coefficients}
\label{app2}
We set, for simplicity, channel 7 (see Fig.~\ref{fig:figura_sistema_appendice}) as the lead connected to the reference reservoir. Moreover, we set the relative dynamical phase $k\Delta L=\pi/2$ in order to maximize the effect of the sign flip of $\bf{B}$. Using the Landauer-B\"uttiker formalism\cite{landauer1957,buttiker1986} and the scattering coefficients from Appendix~\ref{app1}, we compute the Onsager coefficients,
\begin{align}
&L_{11}=T\,\int\, \text{d}E\, \left(-\partial_E f\right)\,\tau^2,\cr
&L_{12}=T\,\int\, \text{d}E\, E\,  \left(-\partial_E f\right)\,\tau^2=L_{21},\cr
&L_{22}=T\,\int\, \text{d}E\, E^2\, \left(-\partial_E f\right)\,\tau^2,\cr
&L_{33}=T\,\int\, \text{d}E\, \left(-\partial_E f\right)\,\left[1-\frac{1}{4}\,\cos^2\Phi\,(1+\rho)^2\right],\cr
&L_{34}=T\,\int\, \text{d}E\, E\, \left(-\partial_E f\right)\,\left[1-\frac{1}{4}\,\cos^2\Phi\,(1+\rho)^2\right]=L_{43},\cr
&L_{44}=T\,\int\, \text{d}E\, E^2\, \left(-\partial_E f\right)\,\left[1-\frac{1}{4}\,\cos^2\Phi\,(1+\rho)^2\right],\cr
&L_{13}=T\,\int\, \text{d}E\, \left(-\partial_E f\right)\,\left[-\frac{1}{2}\,\tau^2\,(1+\sin\Phi)\right],\cr
&L_{14}=T\,\int\, \text{d}E\, E\,\left(-\partial_E f\right)\,\left[-\frac{1}{2}\,\tau^2\,(1+\sin\Phi)\right]=L_{23},\cr
&L_{24}=T\,\int\, \text{d}E\, E^2\,\left(-\partial_E f\right)\,\left[-\frac{1}{2}\,\tau^2\,(1+\sin\Phi)\right],\cr
&L_{31}=T\,\int\, \text{d}E\, \left(-\partial_E f\right)\,\left[-\frac{1}{2}\,\tau^2\,(1-\sin\Phi)\right],\cr
&L_{41}=T\,\int\, \text{d}E\, E\,\left(-\partial_E f\right)\,\left[-\frac{1}{2}\,\tau^2\,(1-\sin\Phi)\right]=L_{32},\cr
&L_{42}=T\,\int\, \text{d}E\, E^2\,\left(-\partial_E f\right)\,\left[-\frac{1}{2}\,\tau^2\,(1-\sin\Phi)\right],
\label{eq:onsagers_app}
\end{align}
where $f=\left[\text{exp}\{E/k_BT\}+1\right]^{-1}$ is the equilibrium Fermi distribution at temperature $T$ and $\mu=0$. %
%
%
%
%---------------------------------------------------------------------------------
%\pagebreak

% ---------------------------------------------------------------------------------------------


\begin{thebibliography}{88}
\bibitem{NOTE} 
 The only limitation being not to set all of the $\{X_k^T\}$'s to be null: This configuration in fact only
 admits $\{X_k^\mu\}=0$ $\forall \, k$, hence no currents would flow through the device. 
In this scenario the $\{X^\mu_k\}$'s become explicit functions of the temperatures $\{X_k^T\}$ and of the Onsager coefficients $L_{ij}$.
\bibitem{note_reversibility}
Note, however, that in general the entropy production rate is nonzero, 
due to the irreversible component of the particle current.
\bibitem{note_nointeracting}
We have used the fact that, due to the symmetries of the scattering matrix, 
the Onsager matrix is block diagonal, and therefore $L_{21}=L_{12}$
and $L_{43}=L_{34}$.
\end{thebibliography}
\end{document}